\documentclass{aa}
\usepackage{graphics,epsfig}

\newcommand{\grs}    {GRS 1915+105}
\def\simless{\lower.5ex\hbox{\ltsima}}            
\def\simmore{\lower.5ex\hbox{\gtsima}}            

\def\msun{{\rm M}_\odot}

\begin{document}

\title{The aperiodic variability of Cyg X-1 and GRS 1915 +105 at very low
frequencies}

\subtitle{}

\author
{P. Reig \inst{1,2}, I. Papadakis\inst{1,2}, N.D. Kylafis\inst{1,2}
}

\institute{
Foundation for Research and Technology-Hellas, GR-711 10 Heraklion, Crete, 
Greece
\and Physics Department, University of Crete, P.O. Box 2208, GR-710 03 
Heraklion, Crete, Greece
}

\authorrunning{Reig et al.}
\titlerunning{Variability of Cyg X-1 and GRS 1915 +105 at very low frequencies}

\offprints{pablo@physics.uoc.gr}

\date{Accepted \\
Received : \\
}

\abstract{ 
We have carried out a timing analysis of Cyg X--1 and \grs\ using more
than 5 years worth of data from the {\em All Sky Monitor} (ASM) on board
the {\em Rossi X-ray Timing Explorer} (RXTE). We have obtained for the
first time power density spectra, colour-colour and colour-intensity
plots, cross-correlation functions and phase-lag diagrams to investigate
the variability of the sources at frequencies $< 10^{-5}$ Hz. We find that
the power spectra are not flat but consistent with a power-law of index
$\sim -1$. This is the same slope as that found in the power spectra of
Cyg X--1 during the soft state. In fact, the power spectrum of Cyg X--1 in
the frequency range $10^{-7}-10^{-5}$ Hz appears as a continuation of the
$10^{-3}-10$ Hz power spectrum during the soft state. Significant
variability, expressed as the fractional $rms$ in the frequency range $2
\times 10^{-7}-1 \times 10^{-5}$ Hz, is detected at a level of 27\% and
21\% for \grs\ and Cyg X--1, respectively.  Our results  confirm previous
suggestions that the innermost regions of the accretion disc, where the
X-ray photons are produced, are affected by variations occurring at larger
radii from the black hole, presumably due to changes in the mass accretion
rate. We also observe strong spectral changes: in Cyg X--1 the spectrum
softens as the flux increases when it is in the low/hard state, and
hardens with increasing flux in the high/soft state. \grs\ follows the
same trend as Cyg X--1 in the soft state, but the hardness ratios show a
larger amplitude of variation. The cross-correlation shows that the
variations occur simultaneously in all energy bands. However, the
cross-correlation functions are asymmetric toward negative lags. The phase
spectrum also reveals negative lags at periods larger than a few days,
suggesting that the long term variations in the soft energy band are
delayed with respect to the harder bands. 
}

\maketitle

\section{Introduction}

Accreting black holes are currently among the most intensively
studied objects in high-energy astrophysics. They are the sources of
spectacular phenomena such as X/gamma-ray outbursts and relativistic jets
and constitute important laboratories where the physics of accretion
can be studied. The {\em Rossi X-ray Timing Explorer} (RXTE) has made
unique contributions to our understanding of these extreme objects thanks
to its high timing capability, allowing the study of the X-ray variability
on the dynamical time scales of neutron stars and black holes. While the
frequency range between 10$^{-4}$--10$^3$ Hz is now well studied, i.e.
time scales  from milliseconds to hours, the characteristics of the
variability at longer time scales ($\nu < 10^{-4}$ Hz) are much less
constrained. The study of the variability at such low frequencies is of
great importance in the modeling of accretion discs since, as
current models assume, it provides information on the outermost parts
of the disc where changes are expected to occur on the viscous time scale.

Because of its brightness and the persistent nature of the high-energy
radiation, Cyg X--1 is one of the best studied Galactic X-ray sources. 
It consists of a compact object orbiting the $V=8.9$ O9.7Iab blue
supergiant HDE 226868, with an orbital period of 5.6 days. The mass of the
compact companion has been estimated to be $\sim 10 \msun$ (Herrero et al.
1995), hence lying above the upper mass limit for neutron stars. For a
summary of the properties of the two individual components and the orbital
parameters of the system see e.g. Nowak et al. (1999).  Its X-ray
properties, such as the transient appearance of an ultra-soft component,
the hard power-law spectrum at high energies, flickering, spectral
transitions or low-frequency QPO have been considered as canonical
signatures of black-hole candidacy. Especially important are the source
state transitions since they cover a large range of X-ray flux and energy
spectral evolution. Although up to five different  spectral states have
been identified in black-hole  systems,  the  two main ones are  the  {\em
low/hard}  and the {\em  high/soft}  states.  The low and  high  terms
refer to the flux level in the 1--10 keV range,  whereas  the terms hard
and soft refer to the absence or  presence  of a  multicolour  blackbody
component  in the same range, respectively. The shape of the power density
spectrum (PDS) in the range $10^{-3}-100$ Hz of Cyg X--1 is well
established.  In the low/hard state the PDS is flat from 0.002--0.2 Hz, it
then steepens to a power-law with index $\sim -1$ up to a few Hz and
steepens again to a slope of $\sim -2$ up to a few tens of Hz (e.g. Nowak
et al. 1999). Above these frequencies Cyg X--1 shows another steepening at
frequency $\sim 40-80$ Hz with the power-law changing to $\sim -2.3$ and a
fractional $rms$ amplitude between 100 and 400 Hz at the level of $\sim 3$\%
(Revnivtsev et al. 2000). In the soft state a single power-law with index
$\sim$ --1 fits the spectrum from 10$^{-4}$ to 15--20 Hz, around which it
changes to a slope of $\sim -2$ (Churazov et al. 2001; Revnivtsev et al.
2000).

\grs\ is a bright and highly variable X-ray source which is believed to
harbour a black hole on the basis of its spectral and temporal
similarities with other dynamically proven black-hole systems. Although
initially it was considered a transient X-ray source it has never
completely switched off since RXTE established continuous coverage in
February 1996. It was the first Galactic object to exhibit superluminal
radio jets (Mirabel \& Rodr\'{\i}guez 1994).  Recent near-infrared
spectroscopic observations by Greiner et al. (2001) have identified a
K-MIII star as the optical counterpart to \grs. Thus, \grs\, unlike
Cyg~X--1, belongs to the class of low-mass X-ray black-hole binaries.
However, no binary mass function or orbital period are known. The form of
the power spectrum of \grs\ is highly variable and correlates with
spectral and luminosity changes (Trudolyubov et al. 1999). \grs\ 
generally displays two spectral states: during the low luminosity
state, also called {\em plateau} state (Dhawan et al. 2000) or state $C$
(Belloni et al. 2000) the power spectrum, in the 0.01--50 Hz frequency
range, is characterised by a strong band-limited noise component (broken
power-law) and one QPO peak (possibly with some harmonics). The continuum
is flat below the QPO peak with power-law index $\sim$ 0.1--0.3 and
steepens above the QPO to slope 2.5-3.0 (e.g. Reig et al. 2000). Most of
the time \grs\ finds itself in a {\em flare} state, characterised by large
X-ray flux variations. In this state the lower part of the power spectrum
($\nu < 1$ Hz) shows a notable rise approaching a simple power-law (e.g.
Trudolyunov et al. 1999). A collection of power density spectra of \grs\
in different states can be found in Morgan et al. (1997).

Previous studies of the long-term variability of Cyg X--1 have
concentrated on the search for periodic signals (Priedhorsky et al. 1983;
Brocksopp et al. 1999; Kitamoto et al. 2000). In this work we carry out a
timing analysis of the aperiodic variability by obtaining power-density
spectra, colour-colour and colour-intensity diagrams, cross-correlation
and phase-lag diagrams.

        \begin{figure}
    \begin{center}
    \leavevmode
\epsfig{file=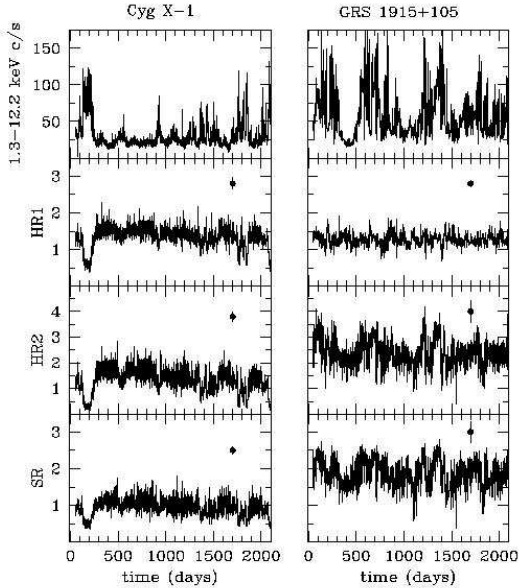, width=8.0cm, bbllx=40pt, bblly=150pt,
  bburx=565pt, bbury=715pt, clip=}
 \end{center}              
        \caption{Light curves and the evolution of the harness ratios for
the entire period. The hardness ratios are defined as HR1=[C]/[B],
HR2=[C]/[A] and SR=[B]/[A], where [A]=1.3-3.0 keV, [B]=3.0-5.0 keV and
[C]=5.0-12.2 keV. Time 0 corresponds to JD 2,450,139.1}
        \label{ctime}
        \end{figure}

\section{Data analysis and results}

The light curves in the different energy ranges, which cover 5.7 years of
data, from February 1996  to October 2001 (JD 2,450,139--2,452,200),  were
retrieved from the {\em All Sky Monitor} (ASM) Definitive Products
Database. The ASM consists of three wide-angle ($6^{\circ} \times
90^{\circ}$) shadow cameras (SSC1--3) equipped with position-sensitive
Xenon proportional counters with a total collecting area of 90 cm$^2$. The
ASM scans $\sim$ 80\% of the sky every $\sim$ 90 minutes in a series of
dwells of about 90 s each. Any given X-ray source is observed in about
5--10 dwells every day. The ASM is sensitive to X-rays in the energy band
1.3-12.2 keV, but also provides a nearly continuous monitoring of the
X-ray sky in three sub-bands: 1.3-3.0 keV (band A), 3.0-5.0 keV (band B)
and 5.0-12.2 keV (band C). We made use of data from SSC2 only since it has
no gain evolution and hence no shifts in channel boundaries. For more
information on the ASM see Levine et al. (1996).

\subsection{Light curves and hardness ratios}

Possible gaps due to detector failure or lack of data were filled by
linear interpolation, adding appropriate random noise. The gaps are
randomly distributed over the whole light curves. For the 1-day rebinned
light curves the missing points represent about 9\% for Cyg X--1 and 18\%
for \grs. The average number of missing points per gap is 2. In order to
have a larger frequency coverage in the power spectra, 0.5-day binned light
curves were used. The results are consistent within the errors with those
obtained using 1-day binned light curves.

We produced three hardness ratios by dividing the intensity of the
different bands in the following way: HR1=[C]/[B], HR2=[C]/[A] and
SR=[B]/[A].

Figure~\ref{ctime} shows the overall 1.3--12.2 keV light curve and the
evolution of the hardness ratios throughout the entire period considered
in this work. Both Cyg X--1 and \grs\ are highly variable sources. When
the whole ASM energy range is considered, they both show a fractional $rms$
amplitude of $\sim$54\%. However, X-ray variability in Cyg X--1 is strongly
dependent on energy. While \grs\ is almost equally variable in all three
energy ranges, with a slight increase of the fractional $rms$ toward harder
energies ([A]: 48\%, [B]: 55\%, [C]: 57\%), hard X-rays in Cyg X--1 are
three times less variable than soft X-rays ([A]: 98\%, [B]: 59\%, [C]:
30\%). The errors in the fractional $rms$ are $\le$ 1\%.

The average values of the hardness ratios are $<{\rm SR}>=0.9\pm0.2$,
$<{\rm HR1}>=1.3\pm0.2$ and $<{\rm HR2}>=1.3\pm0.4$ for Cyg X--1 and
$<{\rm SR}>=1.9\pm0.3$, $<{\rm HR1}>=1.3\pm0.1$ and $<{\rm
HR2}>=2.4\pm0.5$ for \grs. That is, on average, \grs\ shows a much harder
spectrum.

        \begin{figure}
    \begin{center}
    \leavevmode
\epsfig{file=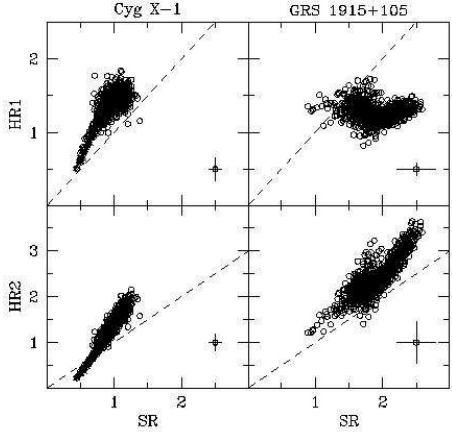, width=8.0cm, bbllx=30pt, bblly=230pt,
  bburx=555pt, bbury=670pt, clip=}
 \end{center}              
        \caption{Colour-colour diagrams. Hardness ratios are defined as in
Fig.~\ref{ctime}.}
        \label{ccd}
        \end{figure}
        \begin{figure}
    \begin{center}
    \leavevmode
\epsfig{file=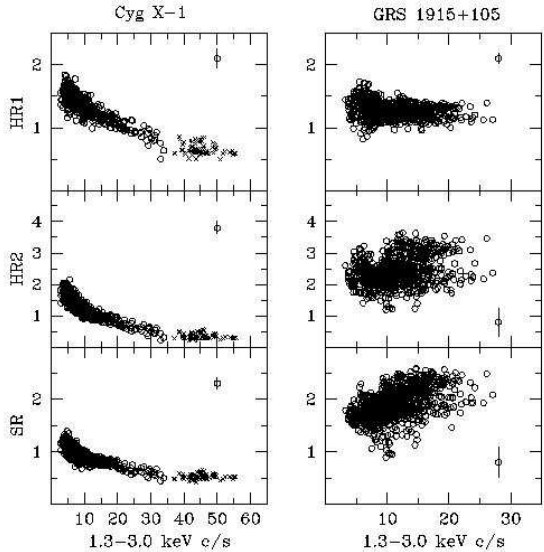, width=8.0cm, bbllx=65pt, bblly=170pt,
  bburx=555pt, bbury=670pt, clip=}
 \end{center}              
        \caption{Hardness ratios as a function of the soft (1.3--3.0
keV) intensity.}
        \label{crate1}
        \end{figure}
        \begin{figure}
    \begin{center}
    \leavevmode
\epsfig{file=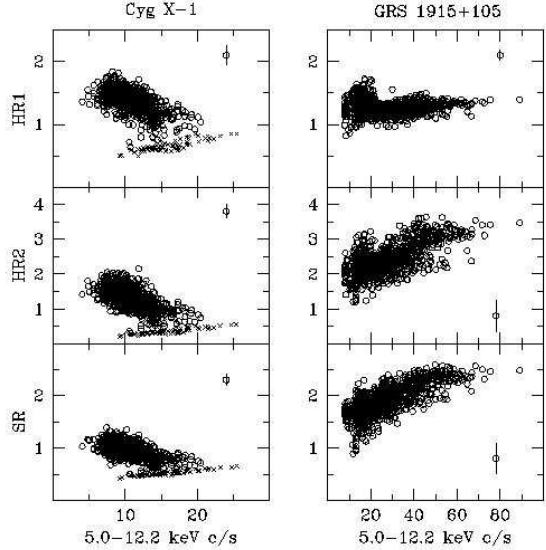, width=8.0cm, bbllx=65pt, bblly=170pt,
  bburx=555pt, bbury=670pt, clip=}
 \end{center}              
        \caption{Hardness ratios as a function of the hard (5.0--12.2
keV) intensity.}
        \label{crate3}
        \end{figure}

\subsection{Colour-colour and colour-intensity diagrams}

Colour-colour diagrams (CCD, Fig.~\ref{ccd}) and colour-intensity diagrams
(CID, Fig.~\ref{crate1} and Fig.~\ref{crate3}) were obtained by plotting
the hard colours HR1 and HR2 versus the soft colour SR and versus the soft
(band [A]) and hard (band [C]) intensities, respectively. For the sake of
clarity the individual errors of each point have been omitted. Instead,
the mean error is plotted.  Likewise, the data were rebinned so that
one point represents a time span of two days.

The canonical low/hard and high/soft spectral states seen in black-hole
systems show up in the CCD of Cyg X--1. During periods of high count rate
the spectra become softer and the source moves to the left bottom part of
the CCD (crosses), i.e. the regions enclosed by [HR2 $<$ 1, SR $<$ 0.8]
and [HR1 $<$ 1, SR $<$ 0.8]. However, Cyg X--1 spends most of its life in
the low/hard state as indicated by the high density of points at values of
the harness ratios larger than 1.  In contrast, in \grs\ these regions are
not populated at all.

The low/hard and high/soft states are clearly separated in the hardness
ratio versus hard intensity diagram of Cyg X--1 (Fig.~\ref{crate3}).  In
the hard state, as the intensity increases the spectrum  becomes softer,
whereas in the soft state the opposite behaviour is observed, as the
intensity increases the spectrum becomes harder. In \grs\ the changes are
similar to those of Cyg X--1 in the soft state. The differences are the
higher values and the larger amplitude of the hardness ratios. \grs\ finds
itself most of the time in a sort of very high state (Belloni et al.
2000).

        \begin{figure}
    \begin{center}
    \leavevmode
\epsfig{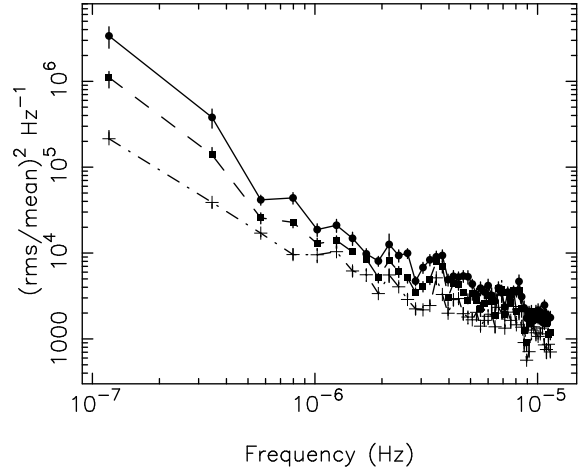}
 \end{center}              
        \caption{Power density spectra of Cyg X-1 at three different
energy ranges: 1.3-3.0 keV (circles), 3.0-5.0 keV (squares) 
and 5.0-12.2 keV (crosses).}
        \label{pdscyg}
        \end{figure}
        \begin{figure}
    \begin{center}
    \leavevmode
\epsfig{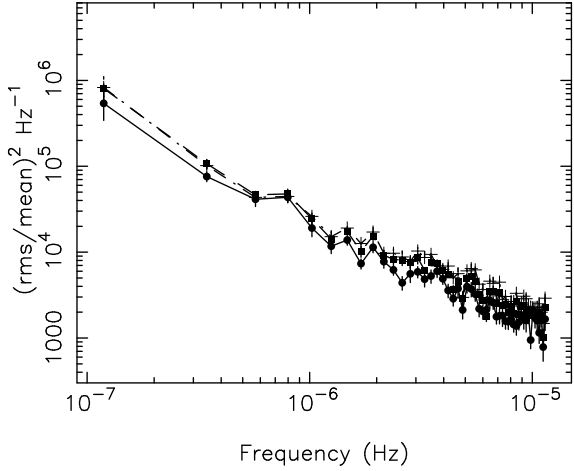}
 \end{center}              
        \caption{Power density spectra of GRS 1915+105 at three different
energy ranges: 1.3-3.0 keV (circles), 3.0-5.0 keV (squares) 
and 5.0-12.2 keV (crosses).}
        \label{pds1915}
        \end{figure}

\subsection{Power density spectra}

In order to investigate the source variability we divided the light curves
into 1024 day segments and computed the power-density spectra (PDS) by
taking the Fast Fourier Transform of each segment. The resulting PDS
were averaged together and rebinned in frequency to have at least 40
points per bin. The  final PDS in each energy band are shown in
Fig.~\ref{pdscyg} and Fig.~\ref{pds1915}. The PDS were normalized such that
their integral gives the squared fractional $rms$ variability (Belloni \&
Hasinger 1990).

The power spectra are characterised by an approximate power-law red-noise
component with no sign of flattening at low frequencies. Although a simple
power-law function did not provide a good fit to the PDS -- the reduced
$\chi^2$ was 1.9 and 2.8 for \grs\ and Cyg X--1, respectively -- it allows
direct comparison of our results with those found at higher frequencies
(see below). The best-fit power-law index and variability (given as the
fractional $rms$ in the frequency range $2 \times 10^{-7}$--$1 \times
10^{-5}$) for the 1.3--12.2 keV PDS was $\alpha=-1.12\pm0.04$, $rms$=27\%
and $\alpha=-0.93\pm0.05$, $rms$=21\% for \grs\ and Cyg X--1, respectively.
While the slope does not show any dependence on energy in either system,
the $rms$ is much more dependent on energy in Cyg X--1 than in \grs\ (see
discussion).

        \begin{figure}
    \begin{center}
    \leavevmode
\epsfig{file=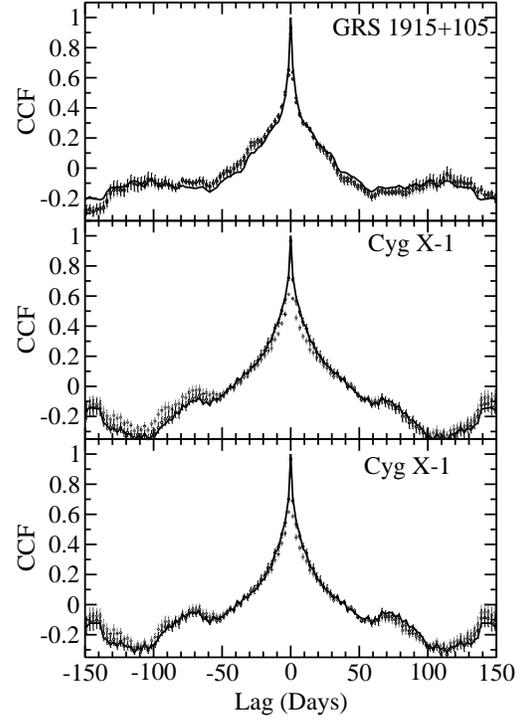, width=7.0cm, bbllx=20pt, bblly=30pt,
  bburx=555pt, bbury=770pt, clip=}
 \end{center}              
        \caption{Cross-correlation functions for the light curves in the
bands [A,B] (open circles) and [A,C] (grey/filled circles). The solid
lines represent
the auto-correlation function of the band [A] light curve. In all cases
the band [A] light curve is taken as the reference light curve, meaning
that CCF peaks at positive lags correspond to the soft X-ray variations
leading the harder ones. The bottom panel shows a plot of the CCF of Cyg
X-1 when we ignore the first 200 days (see Section 2.4).}
        \label{ccrf}
        \end{figure}
        \begin{figure}
    \begin{center}
    \leavevmode
\epsfig{file=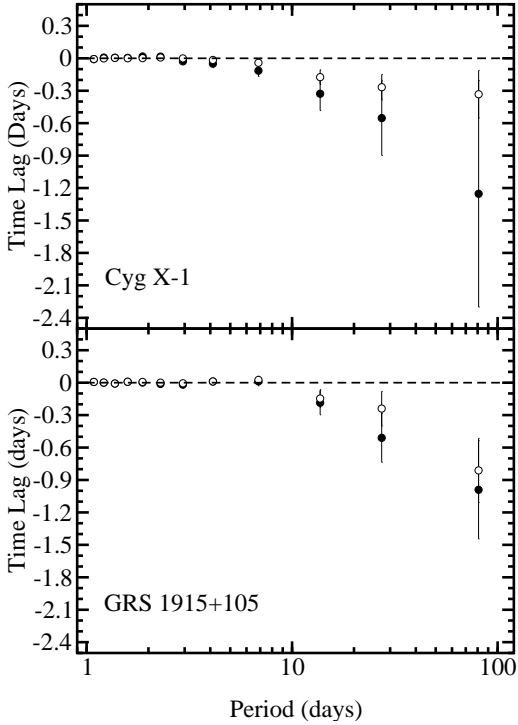, width=7.0cm, bbllx=10pt, bblly=20pt,
  bburx=550pt, bbury=770pt, clip=}
 \end{center}              
        \caption{Time lags versus Fourier period for the phase spectrum of
the band [A] versus band [B] light curves (open circles) and of the band
[A] versus band [C] light curves (filled circles). For both sources, the
time lags are consistent with zero up to a period of $\sim$ a few days,
and then they become negative, decreasing with increasing period.}
        \label{crsp}
        \end{figure}

\subsection{Cross-correlation functions and phase spectra}

In order to study the correlation between the variations in the
different energy bands we computed the cross-correlation function (CCF)
and the phase spectrum between the light curves of the two sources in the
three energy bands.

First, we divided all light curves into 10 parts, 200 days long each.
For each part, we computed the CCF using the following equation,

\[
\hat{\rho}_{sh}(k)=\frac{1}{N\hat{\sigma}_{s}\hat{\sigma}_{h}}
\sum_{t}[x_{s}(t)-\bar{x}_{s}][x_{h}(t+k)-\bar{x}_{h}],
\]

\noindent where $x_{s}(t)$ and $x_{h}(t)$ denote each part of the observed
light curves, $\bar{x}_{s}$ and $\bar{x}_{h}$ denote their respective
means, and $N$ is the number of points in them. The $\hat{\sigma}_{s},
\hat{\sigma}_{h}$ in the above equation are the source standard
deviations, i.e. the square root of the variance of each part minus the
variance that is introduced by the experimental Poisson noise (which we
approximate with the average square error of the points in the respective
part). The summation goes from $t=1$ to $(N-k)$, $k>0$, and from $t=(1-k)$
to $N$, $k<0$. The resulting CCFs were averaged together and rebinned
in order to have 50 points per bin.

In all cases, we chose the lower-energy (i.e. band [A])  light curve as
reference. Therefore, the presence of a CCF peak at positive lags would
mean that the variations in the higher energy emission follow after those
in the softer band. Our results are shown in Fig.~\ref{ccrf}. The solid
lines in this Figure show the auto-correlation functions of the band [A]
light curves. The [A,B] and [A,C] CCFs are shown with the grey/filled and
open circles respectively. The CCFs are peaked at zero lag, with the CCF
peak being larger than $\sim 0.7$ in all cases. This suggests that the
variations in all bands are highly correlated, with no detectable delays
between the different energy bands.

While the CCF peaks are clearly very close to zero lag, the CCFs are not
symmetric. The asymmetry is toward negative lags and it is stronger in the
case of the Cyg X-1 [A,C] CCF. In fact, at lags $< -50$ days, the band [A]
light curve is better correlated with the band [C] light curve than it is
with itself (Fig.~\ref{ccrf}).  This result suggests that the low
frequency component in the harder band leads the respective component in
the softer one.  In order to investigate further the asymmetry in the
CCFs of Cyg X-1, we re-computed them ignoring the first 200 days of the
lightcurves, when the source was in its soft/high state. The results are
shown in Fig.~\ref{ccrf} (bottom panel). The [A,C] CCF is still
assymetric toward negative lags, but this effect is not as pronounced as
before. In \grs, the [A,B] and [A,C] CCFs have similar amplitudes, and
are slightly asymmetric toward negative lags. At lags $<-20$ days the CCF
is larger than the CCF at lags $>20$ days.

We also computed the phase spectrum between light curves at different
energy bands. We first computed the cross spectrum,
$\hat{P}_{sh}(\nu_{j})=X^{\ast}_{s}(\nu_{j})X_{h}(\nu_{j})$, where
$X_{s}(\nu_{j})$ and $X_{h}(\nu_{j})$ are the complex Fourier coefficients
for the two light curves at frequency $\nu_{j}$, and $X_{s}^{\ast}$ is the
complex conjugate of $X_{s}$. We calculated an average cross vector,
$\hat{P}_{sh}(\nu)$, by averaging the real and imaginary parts into bins
containing 200 points  (except the three longest period points for
which we used a bin size of 100 and 50), and then we found the final
value of phase versus frequency. Finally, we used Eq. (16) from Nowak
et al. (1999) to measure its uncertainty.

Our results are shown in Fig.~\ref{crsp}. This Figure shows the time lags
(i.e. the phase spectrum divided by $2\pi\nu$) as a function of the
Fourier period (i.e. $1/\nu$) for the bands [A,B] (open circles) and [A,C]
(filled circles). In both sources, the time lags at periods up to $\sim
5-10$ days are consistent with zero. Above this period, the lags become
negative, and their amplitude increases with period. This trend suggests
that components at frequency $\nu$ in band [A] (soft band) are delayed
with respect to the same components in bands [B] and [C] (hard bands), in
agreement with the asymmetry that the CCFs show toward negative lags.

        \begin{figure}
    \begin{center}
    \leavevmode
\epsfig{file=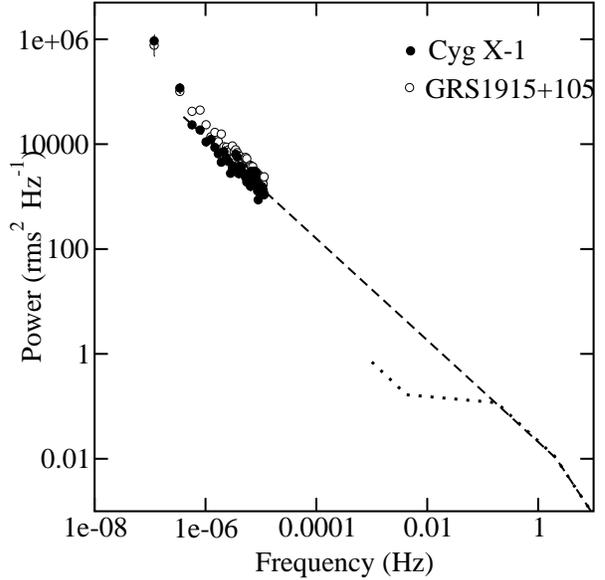, width=8.0cm, bbllx=10pt, bblly=230pt,
  bburx=550pt, bbury=770pt, clip=}
 \end{center}              
        \caption{Comparison of the ASM power density spectrum of Cyg X--1 
and  \grs\ with that observed at higher frequencies. The dashed line represents
the best-fit model to the PDS of Cyg X--1 at higher frequencies when the
source is in the high/soft luminosity state. The dotted
line represents the average PDS of Cyg X--1 in the low/hard state.}
        \label{comp}
        \end{figure}

\section{Discussion}

Black-hole candidates are known to exhibit X-ray variability on a wide
range of time scales from milliseconds to years.  Previous work on the
long-term X-ray variability of Cyg X--1 was carried out by Priedhorsky et
al. (1983), based on data obtained by {\it Vela 5B} and  {\it Ariel 5},
Brocksopp et al. (1999), using data from {\it BATSE}, {\it RXTE}  and
Kitamoto et al. (2000), based on data obtained by {\it GINGA}. They
reported the presence of peaked features in the periodograms of Cyg X--1
corresponding to periods of 294 d (Priedhorsky et al. 1983) and 5.6 d
(orbital period), $\sim$ 150 and 210--230 d (Brocksopp et al. 1999;
Kitamoto et al. 2000). An epoch-folding analysis of the ASM RXTE data
reveals a number of broad-peaked features. However, these features lie on
top of a strong red-noise component, disputing their statistical
significance.

Black-hole X-ray sources display at least two spectral states, {\em
high/soft} and {\em low/hard}, depending on the strength of their soft
($\leq 10$ keV) X-ray emission. The  spectrum at high  luminosities (soft
state) consists of a soft component (normally modeled as a multicolour
blackbody) and a power-law  tail.  This spectrum changes into an 
approximately  single power-law form when the luminosity  decreases (hard
state). The most popular current models invoke  an origin for the soft
X-ray component in an accretion disc and the hard X-ray component via
Comptonization of softer photons in a corona of high-energy electrons.

Cyg X--1 and \grs\ display similarities as well as differences in their
behaviour on long time scales. They both show a similar power density
spectrum in the frequency range $10^{-7}-10^{-5}$ Hz, namely, a strong
red-noise component which can be represented by a power-law function with
index $\sim$ --1. There is no sign of a break in the power law or of the
presence of flat top noise. It is illustrative to compare our
low-frequency  power spectra with those produced at frequencies larger
than $10^{-3}$ Hz, as shown in Fig.~\ref{comp}. The dashed line represents
the best-fit model to the PDS of Cyg X--1 at higher frequencies when the
source is in the high/soft luminosity state (Cui et al. 1997). The dotted
line represents the average PDS of Cyg X--1 in the low/hard state (Nowak
et al. 1999).

The ASM power spectrum of Cyg X--1 in the energy range 1.3-12.2 keV is
consistent with the extrapolation to lower frequencies of its power
spectrum in the range $10^{-3}-1$ Hz during the soft state
(Fig.~\ref{comp}). During the soft state, the disc is believed to extend
all the way to the last stable orbit. The observed variations in this
state are believed to be related to changes in the structure, i.e.
accretion rate, of the disc. Thus, the good agreement between the power
spectra at very low and high frequencies in the soft state and the fact
that significant variability is detected at these very low frequencies
(fractional $rms$ of 26\% in $2 \times 10^{-7}-1 \times 10^{-5}$ Hz) imply
that we are witnessing global changes in the structure of the accretion
disc. The similarity, both in amplitude and shape, between the ASM power
spectra of \grs\ and Cyg X--1 (unlike the high-frequency power spectra)
suggests that the long term variations in \grs\ are also caused by global
changes in the disc structure.

These changes must occur at large radii and propagate on the viscous
diffusion time scale. An order of magnitude estimate of the typical length
scales associated with the low-frequency variability rules out the corona
as the origin of this variability.  Indeed, the viscous time scale is
given by (Frank et al. 1992)

\begin{equation}
t_{visc} \sim 3 \times 10^5 \alpha^{-4/5} \dot{M}^{-3/10}_{16} M^{1/4}_{bh}
R^{5/4}_{10} \, \, \, \, {\rm s}
\end{equation}

\noindent where $\alpha$ is the viscosity parameter, $\dot{M}_{16}$ is the
mass-accretion rate in units of 10$^{16}$ g s$^{-1}$, $M_{bh}$ is the mass
of the black hole in units of the solar mass and $R_{10}$ is the typical
length scale for surface density changes in the disc in units of 10$^{10}$
cm.  Generally accepted values of the size of the Comptonizing corona,
$R_{cor}$, range from 50--100 Schwarzschild radii
($R_{sch}=2GM_{bh}/c^2$). Assuming typical values $\alpha=0.01-0.1$,
$\dot{M}_{16}=1$, $M_{bh}=10$ and $R_{cor}=100 \, R_{sch}$ we find
$t_{visc}\approx 0.2-1$ day, clearly too short to explain the variability
observed in the ASM power spectrum. The variability at the very low
frequencies investigated here requires disc sizes of the order of 2500
Schwarzschild radii.

These results confirm the idea, suggested by Lyubarskii (1997) and
developed by Churazov et al. (2001), that X-ray variability is caused by
instabilities occurring in the accretion disc at very long distances from
the black hole and then propagating into the innermost regions, where most
of the gravitational energy is released and the X-rays are produced.

The differences between Cyg X--1 and \grs\ regard mainly 
the spectral behaviour of the systems and can be summarised as follows:

\begin{itemize}

\item [i)]  The average values of the hardness ratios are higher in \grs\
than in Cyg X--1, indicating that its spectrum is harder. 

\item [ii)] During the hard state the spectrum of Cyg X--1 softens as the hard
count rate increases. During the soft state the opposite trend occurs, i.e
the spectrum is harder when the count rate is high.  In \grs\ only a
positive correlation is seen, namely, the spectrum becomes harder as the
count rate increases.

\item [iii)] Whereas Cyg X--1 exhibits a strong dependence of variability
on energy, the variability of  \grs\ is much less sensitive to energy. In
\grs, the fractional $rms$ in the $2 \times 10^{-7}-1 \times 10^{-5}$ Hz
range stays at a level of 27\%, perhaps showing a slight increase from
band [A] (24\%) to band [C] (28\%). In Cyg X--1 the $rms$ goes from 27\%
in band [A] to 17\% in band [C].

\end{itemize}

These differences can be explained by the limited energy range of the ASM
(1.3--12.2 keV) and by the fact that, different components affect the
X-ray emission in this band in the two sources differently. The long-term
hardness ratio variations imply that the spectrum of Cyg X--1 lacks, most
of the time, a soft component, whereas such a component seems to be an
almost permanent feature in \grs.  This idea is borne out by the
results of the fits of broad-band spectra  to the disc blackbody plus
power-law component model. In \grs\ the soft component, i.e. the
multi-colour blackbody, although variable in strength is always present,
with colour temperature of the disc at the inner radius $\ge$ 0.7 keV
(Muno et al. 1999; Trudolyubov et al. 1999). In contrast, the
characteristic colour temperature of the  soft disc blackbody component in
Cyg X--1 remains $\le$ 0.3 keV (Dotani et al 1997, Frontera et al. 2001).

The trends in the CID of \grs\ are similar to those of the soft state of
Cyg X--1. The main difference is, however, the fact that the
hardness-ratio variations are of larger amplitude in \grs. These
variations could be caused  by changes of the temperature of the
multicolour blackbody component. As the temperature increases, the band
[C] count rate increases and the spectrum becomes harder (i.e. the HR2 and
SR values increase - Fig.~\ref{crate3}). The hardening of the spectrum as
a function of the band [A] count rate (Fig.~\ref{crate1}) also suggests
that a temperature increase is associated with an increase in the
amplitude of the blackbody component. The small amplitude spectral
variations in Cyg X--1, when in the soft state,  could be explained 
assuming that the temperature of the blackbody component stays roughly
constant, as it has probably reached its highest value. At this stage, any
significant temperature variations are toward the direction of smaller
temperature values, causing the object to go back to the hard state. The
flux variations could be caused by blackbody amplitude variations which
under approximately constant temperature would produce small spectral
variations (as observed).

The much larger average values of the hardness ratios in \grs\ 
(especially HR2) also support the idea of a blackbody component being
present most of the time and having a higher temperature (on average) than
in Cyg X--~1. In \grs\ the long-term variations result from changes in the
rather hot soft component (hot enough to affect all three energy bands).
In Cyg X--1, bands [B] and [C] are probably significantly affected by
changes in the hard power-law component as well. Therefore, in Cyg X--1,
the X-ray variability is the result of variations in the soft as well as
in the hard component, explaining the strong dependence of $rms$ on energy.
In \grs\ none of the three energy bands considered here is strongly
affected by the power-law component. The X-ray variability reflects
changes in the soft component, explaining the similarity of $rms$ values in
all three energy bands. 

Finally, the cross-correlation analysis results show that the variations
in the various energy bands occur simultaneously in both objects. The
maximum correlation appears at lag zero, and is similar (in amplitude) in
all CCFs except for the band [A,C] CCF of Cyg X-1. This result gives
further support to the idea that in Cyg X-1 we are witnessing the
contribution of two different components (i.e. the soft and the power law)
in these bands. The decrease of the CCF peak is the result of the
flux-dependent spectral changes of the source. Despite the fact that the
maximum CCF peak appears at lag zero, the CCFs are asymmetric toward
negative lags. This result is in agreement with the observed phase lags,
which become negative at periods larger than $\sim$ a few days. This
behaviour is different from what is observed at short time scales, where
the variations at energy bands $> 4$ keV lag behind those at the soft band
(i.e. $<4$ keV) and the phase lags increase positively with period (Nowak
et al. 1999). Both the CCF and the negative lags indicate that the long
time scale variations in the hard band are leading those at the softer
bands in both objects. One possibility is that the small scale variations
affect simultaneously the inner region of the disk which is responsible
for the emission at energies larger than say 1 keV (causing the CCF to
peak at zero lag) while the large amplitude, long-term variations affect
firstly the innermost parts of the disk (which are hotter and emit the
higher energy photons) and then propagate toward larger radii. Perhaps
these variations correspond to significant, structural variations in the
disk (i.e. its ``appearance" or ``disappearance'' of the inner disk) which
propagate from the inner to the outer part.

\section{Conclusion}

The analysis of more than 5 years of continuous monitoring with the ASM
RXTE instrument has revealed significant long-term X-ray variability in
the black-hole systems Cyg X--1 and \grs.  The different long-term
spectral behaviour between Cyg X--1 and \grs\ can be attributed to the
different contribution of the spectral components to the energy range
1.3-12.2 keV. During the low/hard state of Cyg X--1, the hard power-law
component extends well below 10 keV and dominates the spectrum. In
contrast, the soft component accounts for most of the spectrum of \grs\ up
to energies close to 10 keV. The absence of a break in the power spectrum,
and the agreement between the ASM and the high-frequency power spectrum of
Cyg X--1 in the soft state (a power-law with index $\sim -1$) imply that
the long-term variability is likely to be associated with global
structural changes in the surface density of the accretion disc caused by
changes in the mass-accretion rate. The fact that the ASM power spectra of
\grs\ and Cyg X--1 are very similar suggests the same origin for the
long-term variations in \grs\ as well.  Therefore, the disc changes
occurring at large radii must be independent of the properties of the
black hole and the structure of the innermost parts of the accretion disc.

If these perturbations propagate with the viscous time scale, then they
cannot originate in the corona, i.e. in regions close to the black hole.
This is an important result which demonstrates that disc variations at
large radii also manifest themselves in the properties of the X-ray
emission in black-hole binaries. Therefore, accretion disc models should
be able to predict the correct shape and amplitude of the X-ray power
spectra at very low frequencies.  The detection of the break frequency at
which the power spectrum finally flattens will provide important clues
about the size of the accretion disc.

\begin{acknowledgements} 

Data provided by the ASM RXTE teams at MIT and at the RXTE SOF and GOF
at NASA's GSFC. The authors acknowledge partial support via the European
Union Training and  Mobility of Researchers Network Grant
ERBFMRX/CT98/0195. 

\end{acknowledgements}

\end{document}